# A Report on Shape Deformation with a Stretching and Bending Energy

Hui Zhao*    Steven J. Gortler[†]

Harvard University, USA

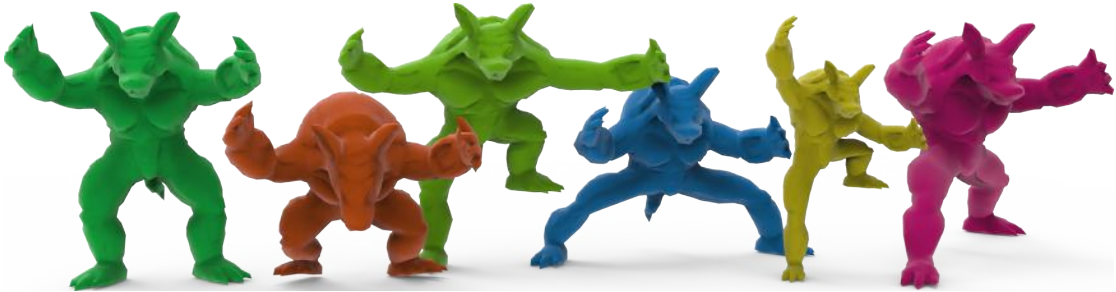

**Figure 1:** *Armadillo Deformation with both Stretching and Bending Energy*

## Abstract

In this report we describe a mesh editing system that we implemented that uses a natural stretching and bending energy defined over smooth surfaces. As such, this energy behaves uniformly under various mesh resolutions. All of the elements of our approach already exist in the literature. We hope that our discussions of these energies helps to shed light on the behaviors of these methods and provides a unified discussion of these methods.

**CR Categories:** I.3.5 [Computer Graphics]: Computational Geometry and Object Modeling—geometric algorithms, languages, and systems

**Keywords:** deformation energy, discrete differential geometry, rotation, digital geometry processing, elasticity, stretching and bending

## 1 Introduction

Interactive shape editing is a fundamental and challenging problem in discrete geometry processing. A central problem in shape editing is to deform an existing mesh so that it satisfies a set of specified constraints, while preserving the local details of the original surface as much as possible [Yu et al. 2004], [Xu et al. 2006], [Botsch and Sorkine 2008], [Sorkine et al. 2004], [Lipman et al. 2004], [Sorkine and Cohen-Or 2004], [Lipman et al. 2005]. In the literature, there are several classes of such deformation methods such as cage-based and lattice-based methods. One of the most popular, intuitive, flexible and predictable interfaces allows the user to simply click-and-drag on mesh vertices. Some energy is then minimized while maintaining these user specified constraints.

Shape deformation should be natural, predictable, physically plausible and aesthetically pleasing. One way to achieve this is using a solid (volumetric) representation of the geometry and then minimize a volumetric "stretching" energy based on elastic mechanics [Chao et al. 2010]. Unfortunately, such methods, are often too expensive for real time editing, and also require a solid representation as input. A simpler alternative is to use a surface-based deformation energy. When dealing with a surface based method, one needs to penalize not only stretching, but also some measure of bending. Otherwise the material will behave very flexibly, like paper or cloth.

One intriguing such method is called ARAP [Sorkine and Alexa 2007]. This method uses an energy that is non-linear, but can be conveniently solved using a "local/global" (alternating minimization) iteration process. For meshes of moderate resolution and moderately uniform tessellation, this energy successfully penalizes both stretching and bending, leading to a very useful editing paradigm.

But as pointed out in [Chao et al. 2010], for this energy, (as well as for a related ARAP Spoke-Rim energy), the resistance to bending emerges from the *size* (and structure) of each vertex neighborhood. As such, it does not behave predictably under different tessellations of the same underlying surface. In the limit, with high tessellations rates, these methods provide almost no resistance to bending at all (See Figure 3 and Appendix B below). This issue has recently been well documented in [Levi and Gotsman 2015].

In this report we describe a mesh editing system that we implemented that uses a natural stretching and bending energy over smooth surfaces. As such, this energy behaves uniformly under various mesh resolutions. All of the elements of this approach already exist in the literature. We hope that our discussions of these energies helps to shed light on the behaviors of these methods and provides a unified discussion of them.

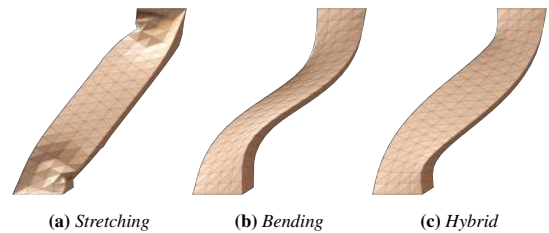

**(a)** *Stretching*  **(b)** *Bending*  **(c)** *Hybrid*

**Figure 2:** *Here we compare the use of a stretching-only, bending-only and the final hybrid energy we implemented. Both stretching and bending needs to be penalized to achieve satisfactory results.*

The stretching energy term that we use, similar to [Liu et al. 2008] and [Chao et al. 2010], is derived from a continuous elastic-energy term. When appropriately discretized, using a triangle based finite element approach, this leads to non-linear, discrete Poisson-like system of equations. The bending energy term that we use, similar to that of [Bouaziz et al. 2014; Au et al. 2005; Au et al. 2006] is derived from a simple measure of mean-curvature distortion. When appropriately approximated and discretized, this leads to non-linear, discrete biPoisson-like system of equations. Alone, this term does not penalize stretching and thus does not result in a satisfactory mesh editing tool. But together, these two energy terms result in a well behaved mesh editing system. (See Figure 2).

---
*e-mail:alanzhaohui@seas.harvard.edu
[†]e-mail:sig@cs.harvard.edu

The energy we use naturally decomposes into a form that can be optimized using a local/global iteration process where the linear system is of the familiar "Poisson plus biPoisson" form. This method is variational and every step is guaranteed to lower an energy form.

## 2 Related Work

The graphics research community has proposed many energy functions to measure shape deformation. Many such proposals start with some ideal goal of measuring the amount of stretching and bending that a surface undergoes as it deforms form its starting state $S$ to a deformed state $S'$. Such an ideal energy should be invariant under rigid motions of $S'$, and should be zero only for when $S$ is related to $S'$ through a rigid motion. This typically results in an energy minimization problem is non-linear on vertex positions.

In a first generation of mesh editing systems, such non-linear energies were considered to be too expensive. This point of view led to a large variety of approaches that attempted to use simpler, quadratic, energies resulting in a single linear system. These are comprehensively surveyed in [Botsch and Sorkine 2008]. Such linear methods include the original Laplacian/Poisson method of [Yu et al. 2004] as well as the biLaplacian/biPoisson type methods: [Lipman et al. 2004] and [Sorkine et al. 2004]. As described by that survey, these linear methods can work well if rotational constraints are explicitly placed by the user, and then propagated using the described algorithms. But when the user only specifies point constraints (dragging some vertices), these linear methods need to somehow infer the rotational components. Algorithms that have been proposed to infer or propagate rotations include [Zayer et al. 2005], [Sorkine et al. 2004] and [Lipman et al. 2005]. As surveyed in [Botsch and Sorkine 2008] this typically leads to unsatisfactory results. As a result of this unsatisfactory behavior, researchers have come to realize that we indeed do need to use a non-quadratic energy for general mesh editing.

The ARAP(As-rigid-as-possible) algorithm of [Sorkine and Alexa 2007] derives a energy that, though non-quadratic, allows for an elegant local/global iteration strategy. In their model, they picture the discrete surface as covered by small overlapping cells. A cell consists a vertex and its one-ring neighbors. The ARAP energy is then defined as the deviations from rigidity of each cell. This method works by simultaneously optimizing over vertex positions as well as rotational variables associated with each cell. For fixed vertex positions, these rotations can be optimally solved for locally at each vertex neighborhood. For fixed rotations, the optimal vertex positions can be found by solving a Poisson-like linear system.

As noted by [Chao et al. 2010], for this deformation energy, as well as for a related ARAP Spoke-Rim energy, the resistance to bending derives from the discrete nature of the overlapping cells. At high tessellation rates, the cells sized become small and the bending resistance vanishes. With uneven triangulations, the bending resistance is spatially variable. This behavior has been documented in [Levi and Gotsman 2015]. As noted in [The CGAL Project 2015], the ARAP Spoke-Rim energy is always positive definite, while the original ARAP energy can be indefinite for some meshes. Figure 3 shows how the ARAP energies can behave sensitively to the tessellation resolution, and compares to the method we implemented. This issue is discussed in more detail in Section B. If figure 4 shows how the original ARAP method can behave very different under irregular triangulations, and compares this to the method we implemented.

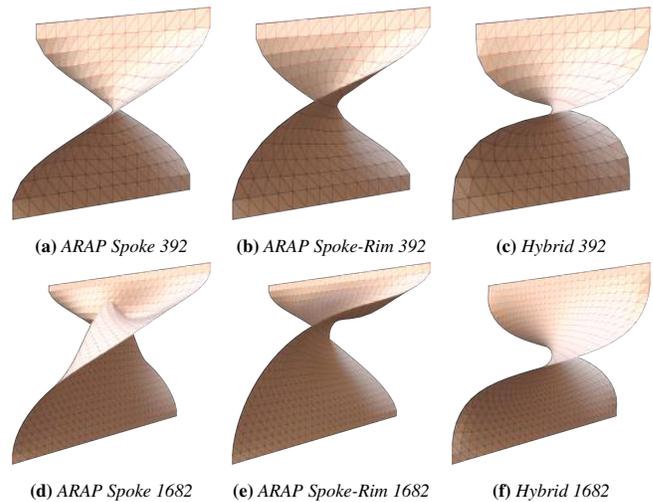

(a) *ARAP Spoke 392*  (b) *ARAP Spoke-Rim 392*  (c) *Hybrid 392*

(d) *ARAP Spoke 1682*  (e) *ARAP Spoke-Rim 1682*  (f) *Hybrid 1682*

**Figure 3:** *Here we twist a regularly triangulated flat sheet. Three algorithms are shown, the original ARAP (spoke) method, the ARAP-Spoke-Rim variant, and the hybrid method we implemented. At low tessellation rates, the three behave similarly. At higher tessellation rates, the two ARAP methods lose their ability to resist bending. The hybrid method we implemented is insensitive to this resolution change.*

Inspired by the elegance of the ARAP method, [Liu et al. 2008] developed a related local/global parameterization algorithm that measured the metric deformation between a mesh surface in 3D and a 2D parameterization. By using a triangle based (instead of cell based) approach, they were able to derive their method as a direct discretization of an intrinsic stretching energy between one mesh surface in 3D and a mesh surface in the plane. Since the target mesh lies in the plane, they do not need to consider bending.

The paper by [Chao et al. 2010], showed how a triangle based (instead of cell based) approach could lead to a direct discretization of an intrinsic continuous stretching energy between two mesh surfaces in 3D. They also apply their ideas to measure stretching for a tetrahedral solid representation. In this context, we note (see Figure 2) that their surface based approach in 3D does not account for bending, and thus results in overly wrinkly behavior. In [Levi and Gotsman 2015], a regularization term is added to the triangle based stretching energy, which they demonstrate leads to very well behaved deformations.

The papers by [Au et al. 2005; Au et al. 2006] describe mesh editing systems based on non-linear biPoisson problems. Although, they do not explicitly say so, as we describe below, these are essentially derivable using a natural mean-curvature energy. Due to some specific choices they made, they needed in [Au et al. 2006] to employ a complicated dualization scheme. In this context, we note (see Figure 2) that their approach does not account for stretching, and thus results in overly rubber-like behavior. In [Au et al. 2006], this problem is noted, and an attempt to ameliorate it is proposed using on an ad-hoc "rescaling" step.

Recently [Bouaziz et al. 2014] derived a new bending energy based on the absolute value of mean curvature. This energy can be easily optimized using a local/global approach, and it is this bending energy that we have chosen to include in our implementation.

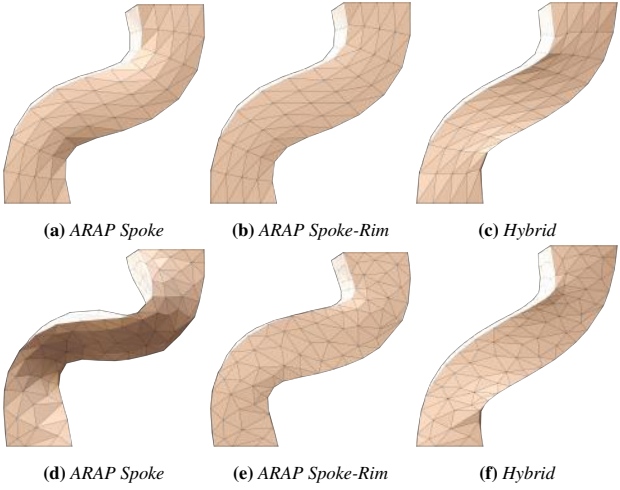

**Figure 4:** *Here we pull on the top face of a rectangular bar. Three algorithms are shown, the original ARAP (spoke) method, the ARAP-Spoke-Rim variant, and the hybrid method we implemented. On the top row, a regular triangulation of the underlying geometry is employed. On the second row, the underlying triangulation is changed. In this case, the ARAP result changes significantly. The ARAP-Spoke-Rim as well as the hybrid method we implemented does not.*

(a) *ARAP Spoke*  (b) *ARAP Spoke-Rim*  (c) *Hybrid*
(d) *ARAP Spoke*  (e) *ARAP Spoke-Rim*  (f) *Hybrid*

Many authors have also proposed other purely discrete, non-linear deformation energies to measure stretching and bending such as [Botsch et al. 2006] and [Fröhlich and Botsch 2011]. These will be beyond the scope of our discussion, as they are not based on the geometry of a smooth underlying surface.

## 3 Energy

### 3.1 Stretching Energy

In this section, we describe the stretching energy between two surfaces which mirrors the approaches of [Liu et al. 2008] and [Chao et al. 2010].

Let $S$ be be an original (say smooth) surface embedded in 3-dimension, and $p$ be a point on $S$. Let $\mathbf{x}(p)$ represent the 3-vector valued coordinate function over $S$. Let the symbol $g$ represent the metric (symmetric $(0,2)$ tensor) of $S$. Let $S'$ be a deformed surface, the image of $S$ under some map $m$. We will describe the geometry of $S'$ using a new 3-vector valued coordinate function $\mathbf{x}'(p)$ "pulled back" to the original $S$. (The deformed surface will also have its own metric $g'$ which we will not directly use in our calculations.)

The derivative (tangential covector) of a scalar function $f$ over $S$ will be denoted as $df$. If $w$ is a tangential covector, then the symbol $|w|_g^2$ will represent the squared-norm of $w$, induced by the metric $g$ over tangential co-vectors on $S$.

For a 3-vector valued function, $\mathbf{f}$: $f^i$ for $i=1,2,3$, the derivative, $d\mathbf{f}$ will simply be a triplet of tangential covectors. If $\mathbf{w}$ is a triplet of tangential covectors: $w^i$ for $i=1,2,3$, then symbol $|\mathbf{w}|_g^2$ will mean $\sum_{i=1}^{3}|w^i|_g^2$.

If $R$ is a 3-by-3 rotation matrix, we can apply it in the obvious way (on the left) to a triplet of scalars or a triplet of tangential covectors.

Finally let $R(p)$ be a rotation matrix field defined at each point in $S$.

We define the following stretching energy

$$E_s(\mathbf{x}', R) := \int_s |d\mathbf{x}'(p) - R(p)d\mathbf{x}(p)|_g^2 \ dA_g(p) \quad (1)$$

By minimizing over the rotation field, we obtain a stretching energy between $S$ and $S'$.

$$E_s(\mathbf{x}') := min_R \ E_s(\mathbf{x}', R) \quad (2)$$

In the computer graphics literature, one often encounters forms of Equation 1 where the Frobenius norm is used. But these matrix coefficients and the Frobenius norm of the matrix are not invariant to parameterization changes, and thus this norm is not generally appropriate. In the discrete case, where an isometric parameterization over a triangle can always be chosen, the Frobenius norm can be used without ambiguity.

We point out, that his energy is slightly different in form from those of [Liu et al. 2008] and [Chao et al. 2010]. Those papers are mostly concerned with mappings from a surface to $\mathbb{R}^2$, or from a volume to $\mathbb{R}^3$. Thus they express their energies roughly as

$$\int_s |dm(p) - R(p)|^2 \ dA_g(p) \quad (3)$$

where $dm$ is the linearization of the mapping $m$ (a $(1,1)$ tensor, which is a square matrix in coordinates) and $|\cdot|^2$ is the appropriate squared norm that takes into account both metrics $g$ and $g'$. In this setting $R$ is a 2-by-2 rotation for surfaces and a 3-by-3 rotation for volumes. In a mesh editing context, we are concerned with a surface being deformed into another surface in $\mathbb{R}^3$ and so it is easiest to look at the two differentials $d\mathbf{x}$ and $d\mathbf{x}'$ (3-by2 matrices in coordinates) of the two immersions of $S$ into $\mathbb{R}^3$. In this setting $R$ is a 3-by-3 rotation.

But the end result is the similar to that of found in [Liu et al. 2008]. As described in the Section A, it can be shown that, $E_s(\mathbf{x}')$ measures the quantity

$$\int_s (\sigma_1(p) - 1)^2 + (\sigma_2(p) - 1)^2 \ dA_g(p) \quad (4)$$

where $\sigma_1(p)$ and $\sigma_2(p)$ are the the maximum and (resp.) minimal stretching ratios of a tangent vector of $S$ at $p$ under the differential mapping $d\mathbf{x}'$ from $S$ to $R^3$. As such, $E_s(\mathbf{x}')$ is one natural way to measure the stretching of a deforming surface.

The energy, $E_s$ is well defined over smooth surfaces, but (using the weak derivative) can also be directly applied when $S$ is a triangular mesh and $S'$ is a deformation obtained by altering the vertex positions of $S$. In this case, the geometry of $S$ is specified with a 3-vector $\mathbf{x}_v$ associated with vertex $v$ of $S$. The deformed $S'$ is specified by associating a 3-vector $\mathbf{x}'_v$ with vertex $v$. As the coordinate functions are linear over each triangle, their derivatives will be constant over each triangle, so in the optimal solution of Equation 2, $R$ will be constant over each triangle. Thus we will have one rotation matrix variable per triangle referred to as $R(t)$.

Applying the discrete gradient derivation from [Pinkall and Polthier 1993] to each triangle and combining terms, the energy of Equation 1 becomes

$$\boxed{E_s(\mathbf{x}', R) = \sum_{he_{vw}} \cot(a_{vw}) \|(\mathbf{x}'_v - \mathbf{x}'_w) - R(t_{vw})(\mathbf{x}_v - \mathbf{x}_w)\|^2}$$

$$(5)$$

where $\|\cdot\|^2$ is the standard 3-vector norm. In above, the symbol $he_{vw}$ represents the half edge from the vertex $v$ to $w$, and we sum over all the half edges of the mesh. The symbol $a_{vw}$ is the angle of the corner opposite to the half edge $he_{vw}$ in its triangle. $R(t_{vw})$ is the $3 \times 3$ (variable) rotation matrix associated with the triangle face associated with the half edge $he_{vw}$.

This energy bears a strong resemblance to the the original ARAP energy, as well as the ARAP-Spoke-Rim variation, but as always, the devil is in the details. The original ARAP energy can be described as

$$E_{arap}(\mathbf{x}', R) := \sum_v \sum_{w \in N(v)} \left[ cot(a_{vw}) + cot(a_{wv}) \right]$$
$$\|(\mathbf{x}'_v - \mathbf{x}'_w) - R(v)(\mathbf{x}_v - \mathbf{x}_w)\|^2$$
$$= \sum_{he_{vw}} \left[ cot(a_{vw}) + cot(a_{wv}) \right]$$
$$\|(\mathbf{x}'_v - \mathbf{x}'_w) - R(v)(\mathbf{x}_v - \mathbf{x}_w)\|^2 \quad (6)$$

Here $R(v)$ is a rotation matrix associated with each vertex of the mesh. The set $N(v)$ are the vertices adjacent to vertex $v$. The ARAP-Spoke-Rim energy is

$$E_{arap-SR}(\mathbf{x}', R) := \sum_v \sum_{(m,n) \in E(v)} \left[ cot(a_{mn}) + cot(a_{nm}) \right]$$
$$\|(\mathbf{x}'_m - \mathbf{x}'_n) - R(v)(\mathbf{x}_m - \mathbf{x}_n)\|^2 \quad (7)$$

The edge set $E(v)$ include all of the edges in the triangles adjacent to $v$.

The energy in Equation 5 is a direct evaluation of the continuous energy of Equation 1, and only penalizes stretching. The energy of Equation 6 (as well as the related ARAP Spoke-Rim energy) penalizes both stretching and bending through its use of overlapping discrete cells. As described in [Chao et al. 2010], the bending penalty drops as the triangulation resolution rises, and thus arises somewhat accidentally from the discretization.

As described in [The CGAL Project 2015], the ARAP energy can be indefinite, and thus unminimimizable for some meshes. In contrast, the ARAP Spoke-Rim energy is always positive semi-definite. Similarly, we see that Equation 1 is clearly positive semi definite and thus so too must be the energy of Equation 5.

### 3.2 Bending Energy

A bending energy is crucial to the control the deformation of surfaces in 3 dimensions.

The stretching energy, described above already implicitly controls the difference in Gauss curvature between $S$ and $S'$, as an isometric deformation must preserve the Gauss curvature. We also note, that for a *compact surface*, a fundamental result [Lawson and Tribuzy 1981], states that there can be at most two smooth immersed surfaces that are isometric and agree everywhere on a non-constant mean curvature field. In this context, we will use a bending energy term that is based on mean-curvature and described by [Bouaziz et al. 2014]. This energy can can be minimized using a biPoisson-like system. A related energy of this type was suggested, for example, in [Wardetzky et al. 2007]. It is also related to methods described by [Au et al. 2005; Au et al. 2006].

We begin by specifying an ideal mean-curvature bending energy of a smooth deformation $m$, mapping an orientable surface $S$ to $S'$. As above, the deformation is described as the vector valued $\mathbf{x}'(p)$ pulled back over $S$. Let $H(p)$ be the mean curvature of $S$ at $p$. Let $H'(p)$ be the scalar mean curvature of $S'$ at $m(p)$. We can then define the mean curvature energy

$$E_m(\mathbf{x}') := \int_s (|H'(p)| - |H(p)|)^2 \ dA_g(p) \quad (8)$$

Here $|\cdot|$ is absolute value. Indeed, it would seem more natural to avoid this absolute value, (so we could penalize convex/concave flips). But, as we will see below, the use of absolute values allows for a simple local/global iteration with guaranteed energy convergence. Additionally, when combined with the stretching energy, and when (due to continuous user interaction) the initial condition is not far from the solution, the results do not appear to suffer from its addition.

Let $\triangle$ be the Laplace-Beltrami operator on $S$ that maps scalar functions to scalar functions, and maps 3-vector valued functions to 3-vector valued functions. Likewise let $\triangle'$ be the Laplace-Beltrami operator on $S'$. Let $\mathbf{n}(p)$ and $\mathbf{n}'(p)$ be the unit vectors of $S$ and $S'$ at $p$ and $m(p)$ respectively. satisfying $|H(p)|\mathbf{n}(p) = \triangle(\mathbf{x})(p)$. and $|H'(p)|\mathbf{n}'(p) = \triangle'(\mathbf{x}')(p)$. These mean curvature normals, $\mathbf{n}(p)$ and $\mathbf{n}'(p)$, may point into or out the surface depending on the sign of the mean curvature.

Then we can write the above energy as

$$E_m(\mathbf{x}') = \int_s \|\triangle'(\mathbf{x}')(p) - |H(p)|\mathbf{n}'(p)\|^2 \ dA_g(p) \quad (9)$$

In order to move towards a local/global setting, we next define

$$E_m(\mathbf{x}', \mathbf{u}') := \int_s \|\triangle'(\mathbf{x}')(p) - |H(p)|\mathbf{u}'(p)\|^2 \ dA_g(p) \quad (10)$$

where $\mathbf{u}'$ is field of unit 3-vectors.

Since $|H(p)|$ is always positive, then for a fixed $\mathbf{x}'$, if we minimize over $\mathbf{u}'(p)$, a field of unit 3-vectors, at each $p$, the optimal $\mathbf{u}'$ must line up with $\mathbf{n}'$. Thus we can see [Bouaziz et al. 2014] that

$$E_m(\mathbf{x}') = min_{\mathbf{u}'} \ E_m(\mathbf{x}', \mathbf{u}') \quad (11)$$

Note that if we attempted this derivation but started with $(H(p) - H'(p))^2$, using signed mean curvatures with no absolute value taken, then we could not, for a fixed $\mathbf{x}'$, and at any point where the signs of $H(p)$ and $H'(p)$ disagree, compute the resulting energy as a minimum over a unit vector $\mathbf{u}'$. Indeed, at such points, it would require flipping $\mathbf{u}'$ from its minimum energy direction, to in fact its maximum energy direction. This is exactly what was done in [Au et al. 2005] which as reported in [Au et al. 2006] leads to "instability". In [Au et al. 2006] they attempt to fix this problem with a more complicated dualization scheme (which cannot work on surfaces with boundary). They report more stability with this dualization scheme, but as they are (implicitly) basing their method on a signed mean curvature, they still cannot claim that their method converges.

When the mapping from $S$ to $S'$ is isometric, then $\triangle = \triangle'$ [Wardetzky et al. 2007]. Since we are in the setting where we will be penalizing stretching as well, we will approximate $\triangle'$ by $\triangle$ to obtain a continuous bending energy defined as

$$E_{cb}(\mathbf{x}', \mathbf{u}') := \int_s \|\triangle(\mathbf{x}')(p) - |H(p)|\mathbf{u}'(p)\|^2 \ dA_g(p) \quad (12)$$

and

$$E_{cb}(\mathbf{x}') := min_{\mathbf{u}'} \; E_{cb}(\mathbf{x}', \mathbf{u}') \tag{13}$$

In the discrete setting, the smooth $\triangle$ operator is not itself well defined, but there are a variety of discrete (e.g. finite difference based) proxies that can be used in its place. Here we will use the well loved (pointwise) discrete cotan Laplacian defined as

$$L(f)_v := \frac{1}{M_v} \sum_{w \in N(v)} \bigl[ \cot(a_{vw}) + \cot(a_{wv}) \bigr] (f_v - f_w) \tag{14}$$

Here $f$ is a discrete function over the mesh defined by a scalar (resp. 3-vector) $f_v$ at each vertex $v$. The result $L(f)_v$ is a scalar (resp. 3-vector) defined at vertex $v$. The quantity $M_v$ represents the area/mass associated with $v$ (perhaps computed using the Voronoi dual).

With this, the discrete bending energy can be written as

$$\boxed{E_{db}(\mathbf{x}', \mathbf{u}') := \sum_v M_v \left\| L(\mathbf{x}')_v - |H_v| \mathbf{u}'_v \right\|^2} \tag{15}$$

and

$$E_{db}(\mathbf{x}') := min_{\mathbf{u}'} \; E_{db}(\mathbf{x}', \mathbf{u}') \tag{16}$$

where $H_v$ is the (pointwise) discrete mean curvature of $S$ at $v$.

To compute $|H_v|$, we just take the norm of the mean curvature vector, $L(\mathbf{x})_v$.

Again, we point out the similarity of this energy with $E_{arap}$, but once again it is different, and the devil is in the details. The energy $E_{db}$ is derived from a discretization of a clearly defined mean-curvature based bending energy, and by construction, it does not penalize stretching in any way.

### 3.3 Combining the two

In order to obtain satisfactory results (see Figure 2), the stretching energy 5 and the bending energy 15 are combined together into a total discrete deformation energy with a controlling parameter $\lambda \in [0, 1]$:

$$E_t(\mathbf{x}', R, \mathbf{u}') := \lambda E_s(\mathbf{x}', R) + (1 - \lambda) E_{db}(\mathbf{x}', \mathbf{u}') \tag{17}$$

We will also refer to this as the *hybrid* energy.

and

$$E_t(\mathbf{x}') := \lambda E_s(\mathbf{x}') + (1-\lambda) E_{db}(\mathbf{x}') = min_{R,\mathbf{u}'} E_t(\mathbf{x}', R, \mathbf{u}') \tag{18}$$

We note that the $E_s$ term is not invariant to uniform scales applied to both $S$ and $S'$ (though we could easily counter this with a global scale factor based on say, the diameter of the model). As a result, the impact of changes to $\lambda$ are not invariant to global scales as well.

Figure 5 demonstrates the deformations of a cactus by four different parameters. We leave this as a parameter for the user.

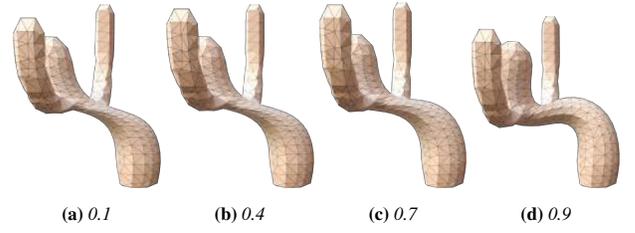

(a) *0.1*    (b) *0.4*    (c) *0.7*    (d) *0.9*

**Figure 5:** *Hybrid mode with four parameters*

## 4 Local Global Iteration

We apply the same local-global methodology of [Sorkine and Alexa 2007; Au et al. 2006; Bouaziz et al. 2014] for mesh deformation, and of [Liu et al. 2008] for mesh parameterization, to solve it.

Our goal is to solve

$$min_{\mathbf{x}'} E_t(\mathbf{x}') \tag{19}$$

subject to the user specified constraints. We reduce this to the problem of minimizing

$$min_{\mathbf{x}', R, \mathbf{u}'} E_t(\mathbf{x}', R, \mathbf{u}') \tag{20}$$

subject to the constraints. This energy is minimized using an alternating minimization strategy.

The local-global approach runs two steps iteratively. In the global phase, we assume that $R$ and $\mathbf{u}'$ are fixed and we solve a linear system over the variable $\mathbf{x}'$. In the local phase, we assume that $\mathbf{x}'$ is fixed and we solve for $R$ and $\mathbf{u}'$. These two steps run iteratively until the energy can not decrease and reaches the minimal value.

### 4.1 Global phase

For a fixed $R$ per triangle and $\mathbf{u}'$ per vertex, the total deformation energy is quadratic in $\mathbf{x}'$. These variables can be optimized by setting the gradient to zero and solving a single linear system.

After combining terms and simplifying, we find that the gradient of the stretching energy with respect to vertex $v$ is

$$\sum_{w \in N(v)} \bigl[ \cot(a_{vw}) + \cot(a_{wv}) \bigr](\mathbf{x}'_v - \mathbf{x}'_w)$$
$$- \sum_{w \in N(v)} \bigl[ \cot(a_{vw}) R(t_{vw}) - \cot(a_{wv}) R(t_{wv}) \bigr](\mathbf{x}_v - \mathbf{x}_w) \tag{21}$$

If we define the 3-vector at vertex $v$

$$\mathbf{b}_v := \sum_{w \in N(v)} \bigl[ \cot(a_{vw}) R(t_{vw}) + \cot(a_{wv}) R(t_{wv}) \bigr](\mathbf{x}_v - \mathbf{x}_w) \tag{22}$$

and then stack these vertically we can describe the gradient of the stretching energy with respect to all of our variables as $ML\mathbf{x}' - \mathbf{b}$ where $M$ is the $n$-by-$n$ diagonal mass matrix. $L$ is the $n$-by-$n$ pointwise Laplacian matrix, $\mathbf{x}'$ and $\mathbf{b}$ are n-vectors of 3-vectors.

We can compute the gradient of the bending energy as $L^t ML\mathbf{x}' - L^t MH\mathbf{u}'$ where $H$ is the n-by-n diagonal matrix of of pointwise absolute values of the mean curvature scalars at each vertex $v$ of $S$, and $\mathbf{u}'$ is an n-vector of unit 3-vectors.

Combining the bending and stretching energy and setting the gradient to zero, we obtain the linear system

$$\boxed{[\lambda ML + (1-\lambda)L^t ML]\mathbf{x}' = \lambda\mathbf{b} + (1-\lambda)L^t MH\mathbf{u}'} \quad (23)$$

As noted in [Botsch and Sorkine 2008] if we pre-multiply on the left by $M^{-1}$, the equation above becomes $[\lambda L + (1-\lambda)LL]\mathbf{x}' = \lambda M^{-1}\mathbf{b} + (1-\lambda)LH\mathbf{u}'$

Our implementation uses positional vertex constraints to drive mesh editing system. When a vertex is constrained, the above linear system is updated in the obvious manner: A constrained $\mathbf{x}_v$ is treated as a constant, the effect of its column in the matrix is pulled over the the right hand side of the equation, and both the row and column associated with $v$ are then removed from the linear system. The matrix of the linear system only needs to be altered, and re-factored when the handle set is changed.

### 4.2 Local phase

In the energy formula, the rotation is defined on the triangle face, and for a fixed $\mathbf{x}'$ this is easy to optimally select at each triangle $t$ independently using a Procrustus computation.

First, we compute a (rank 2) "cross-covariance" matrix $C_t$ [Sorkine and Alexa 2007; Liu et al. 2008] between the two corresponding triangle faces of original and deformed surface by the following formula:

$$C_t = \sum_{i=1}^{3}\cot(\theta_i)\left((\mathbf{x}'_i - \mathbf{x}'_{i+1})(\mathbf{x}_i - \mathbf{x}_{i+1})^T\right) \quad (24)$$

where $i$ sums over three vertices of $t$. Then we compute the SVD $C_t = U^t\Sigma V$. Since $C_t$ is rank 2, we can always choose both $U$ and $V$ have positive determinants without altering $\Sigma$. Finally we compute the optimal rotation for $t$ as $R(t) := U^tV$.

For a fixed $\mathbf{x}'$ it is also easy to compute an optimal unit vector $\mathbf{u}'$ independently at each vertex $v$. For a fixed $\mathbf{x}'$ we compute $L\mathbf{x}'$ and then normalize the resulting 3-vector at each vertex [Bouaziz et al. 2014].

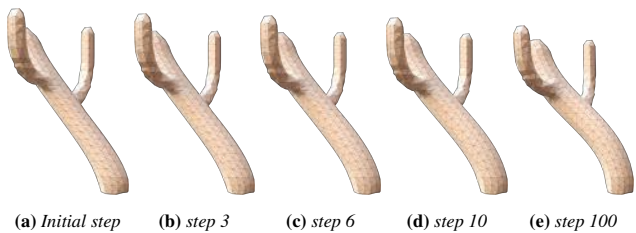

**(a)** *Initial step*  **(b)** *step 3*  **(c)** *step 6*  **(d)** *step 10*  **(e)** *step 100*

**Figure 6:** *Various numbers of iterations are shown.*

In figure 6, the cactus deforms iteratively. It can be seen that the deformation at the step 10 is almost the same with the step 100.

## 5 Experimental Results and Comparison

We have applied this stretching, bending and hybrid algorithms on some specific shape meshes to demonstrate the features and advantages of this method, and compare with a number of other relevant methods.

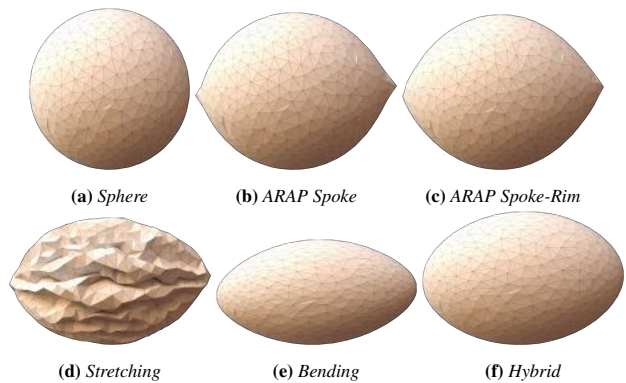

**(a)** *Sphere*  **(b)** *ARAP Spoke*  **(c)** *ARAP Spoke-Rim*

**(d)** *Stretching*  **(e)** *Bending*  **(f)** *Hybrid*

**Figure 7:** *Here we pull on two opposite vertices of a sphere.*

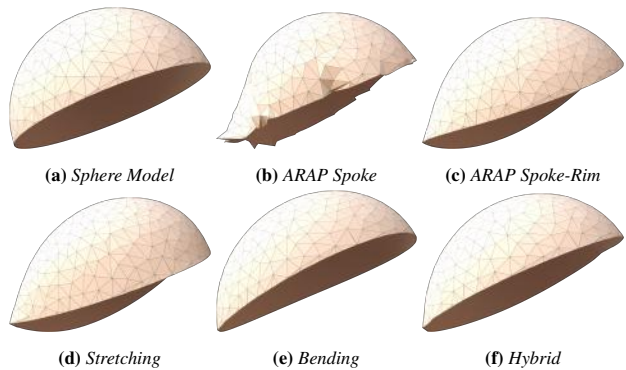

**(a)** *Sphere Model*  **(b)** *ARAP Spoke*  **(c)** *ARAP Spoke-Rim*

**(d)** *Stretching*  **(e)** *Bending*  **(f)** *Hybrid*

**Figure 8:** *Half Sphere Deforms in five algorithms*

In figure 7, we deform a sphere by five algorithms. The original shape is displayed in figure 7a with one point fixed, and another point as moving handle. In these examples, the two ARAP methods maintain the overall spherical shape by concentrating curvature at sharper cusps. The stretching-only result wrinkles up completely. The bending-only result allows for too much surface stretching. The hybrid result demonstrates the best deformation result.

Figure 8b shows a hemisphere being pulled down at one vertex with another vertex fixed. Due to the irregular triangulation, the original ARAP energy is actually negative in this case, the resulting shape is especially poor. In this example, we see the ARAP-Spoke-Rim energy behaving very much like the stretching-only energy.

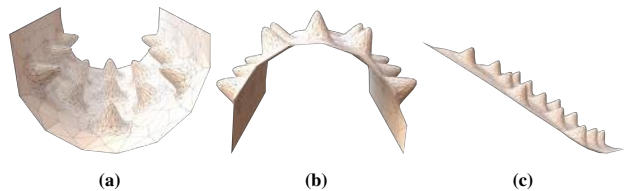

**(a)**  **(b)**  **(c)**

**Figure 9:** *Bump Plane*

In figure 9, a bump plane with some spikes is deformed under the hybrid method. The spikes adjust their orientation automatically according to the positions and orientations of the handle constraints. It shows that this hybrid algorithm propagates the rotations successfully.

In the figure 10, we show that this hybrid algorithm can rotate and twist a bar with large angles successfully. This bending energy is based only on mean curvature, and thus, it does mind when flat regions are turned into saddle regions. This issue can be seen most clearly in in the middle example of this figure.

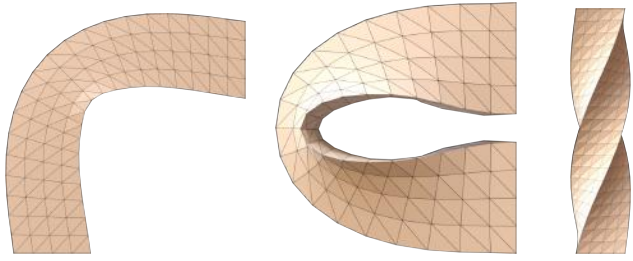

**Figure 10:** *Bar Twist and Rotation by the hybrid mode*

## 6 Conclusion and Future work

Our stretching energy implicitly penalizes changes in Gaussian curvature, as this is determined by the metric. Still, it might be beneficial to have a term that specifically measures changes of Gaussian curvature; for example this might useful to prevent flat regions from turning into minimal-surface/saddle regions, such as seen in Figure 10. Unfortunately, it is not clear to us how to do this, while maintaining the elegant Poisson plus biPoisson formulation, and the local/global optimization structure.

In shape deformation, we only interested in the rest state of the deformation process, and care about the minimization of the deformation energy subject to user defined boundary constraints. However deformation energy can also be used in animating, or simulating surface [Heeren et al. 2014], such as cloth simulation. In future, we would also explore the application of the deformation energy in mesh skeleton, smoothing, shape extrapolation [Kilian et al. 2007], and the segmentation by modal analysis [Huang et al. 2009].

## Acknowledgments

We wish to thank anonymous reviewers for encouragements and thoughtful suggestions. We thank the providers of the free 3D models used in this paper: Armadillo from Stanford Computer Graphics Laboratory; Cactus from AIM@Shape; BumpPlane from Alec Jacobson. We used Mitsuba[Jakob 2010] for rendering images. Our algorithms are implemented on MeshDGP [Zhao 2016] Framework. We also thank Libigl [Jacobson et al. 2016] for reference.

## A  Supplemental: Stretching Energy

In this section we analyze the stretching energy we use and see how it can be understood from the standard point of view of the stretching ratio of tangent vectors under a map. We will work at a single point $p$.

$$E := |d\mathbf{x}'(p) - R^*(p)d\mathbf{x}(p)|_g^2 \qquad (25)$$

where $R^*$ is an optimal rotation.

Since we are dealing with a single point here, we can assume wlog that our surface has a local $(u,v)$ parameterization where at the single point $p$, the 3-vectors

$$\begin{bmatrix} \frac{\partial x^1}{\partial u} \\ \frac{\partial x^2}{\partial u} \\ \frac{\partial x^3}{\partial u} \end{bmatrix}, \begin{bmatrix} \frac{\partial x^1}{\partial v} \\ \frac{\partial x^2}{\partial v} \\ \frac{\partial x^3}{\partial v} \end{bmatrix} \qquad (26)$$

are an orthonormal pair in 3D, and thus in the resulting tangent basis for $TS|_p$, the coordinates of $g$ are represented with the 2-by-2 identity matrix. Expressed in the dual basis, the differential of a function $f$ is has coordinates $[\frac{\partial f}{\partial u}, \frac{\partial f}{\partial v}]$. Thus writing out the energy at $p$ in this dual basis yields

$$E = \left\| \begin{bmatrix} \frac{\partial x'^1}{\partial u} & \frac{\partial x'^1}{\partial v} \\ \frac{\partial x'^2}{\partial u} & \frac{\partial x'^2}{\partial v} \\ \frac{\partial x'^3}{\partial u} & \frac{\partial x'^3}{\partial v} \end{bmatrix} - R^* \begin{bmatrix} \frac{\partial x^1}{\partial u} & \frac{\partial x^1}{\partial v} \\ \frac{\partial x^2}{\partial u} & \frac{\partial x^2}{\partial v} \\ \frac{\partial x^3}{\partial u} & \frac{\partial x^3}{\partial v} \end{bmatrix} \right\|_F^2 \qquad (27)$$

where $\|\cdot\|_F^2$ is the matrix Frobeneous norm.

Let

$$J := \begin{bmatrix} \frac{\partial x'^1}{\partial u} & \frac{\partial x'^1}{\partial v} \\ \frac{\partial x'^2}{\partial u} & \frac{\partial x'^2}{\partial v} \\ \frac{\partial x'^3}{\partial u} & \frac{\partial x'^3}{\partial v} \end{bmatrix} \qquad (28)$$

Next we note that we can always a rotation matrix $R_1$ so that

$$E = \left\| J - R^* R_1 \begin{bmatrix} 1 & 0 \\ 0 & 1 \\ 0 & 0 \end{bmatrix} \right\|_F^2 \qquad (29)$$

Letting $\bar{R}^* := R^* R_1$, we see that $\bar{R}^*$ is the rotation matrix minimizing the energy:

$$E = \left\| J - \bar{R} \begin{bmatrix} 1 & 0 \\ 0 & 1 \\ 0 & 0 \end{bmatrix} \right\|_F^2 \qquad (30)$$

We can use Procrustus analysis to find the form of such an optimal $\bar{R}^*$. We take the covariance matrix

$$\begin{bmatrix} J & 0 \end{bmatrix} = J \begin{bmatrix} I & 0 \end{bmatrix} \qquad (31)$$

And compute its SVD

$$\begin{bmatrix} J & 0 \end{bmatrix} = U^t \begin{bmatrix} \sigma_1 & 0 & 0 \\ 0 & \sigma_2 & 0 \\ 0 & 0 & 0 \end{bmatrix} V \qquad (32)$$

Where $\sigma_1$ and $\sigma_3$ are the singular values of $J$.

And then set the three dimensional rotation as

$$\bar{R}^* := U^t V \quad (33)$$

Plugging this in, to Equation 30, and following reasoning similar to that of [Liu et al. 2008] we can conclude that

$$E = (\sigma_1 - 1)^2 + (\sigma_2 - 2)^2 \quad (34)$$

Finally, to interpret the significance of the $\sigma_i$, let us compute the maximum minimal squared expansion ratios of tangent vectors under the differential of the map from $S$ to $R^3$ defined by $d\mathbf{x}'$. In the $(u, v)$ parameterization, this differential is represented with the matrix $J$. Since the $(u, v)$ parameterization of $S$ is orthonormal at $p$, and since the metric on $R^3$ is the canonical coordinate dot-product, we can calculate these extremal squared stretch values as the eigenvalues of the 2-by-2 matrix $J^t J$, which are $\sigma_1^2$ and $\sigma_2^2$.

## B  Bending Resistance of some energies under refinement

Here we discuss some thought and computer experiments on how various deformation energies behave under the refinement of a triangulation.

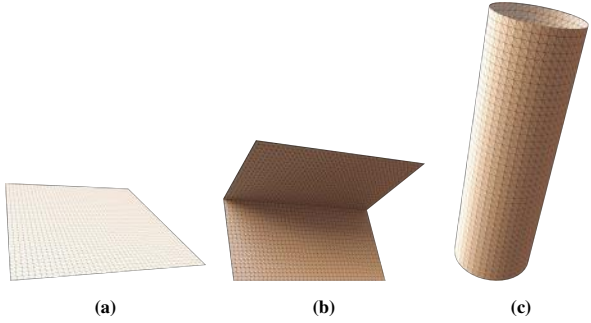

**Figure 11:** *flat sheet, folded sheet, cylinder*

**Case 1**

Consider the case where $S$ is a unit square planar surface, that has been tessellated as a triangulation of an b-by-n grid. Let $S$ be the deformed surface where it has been folded in half along its center line (along a path of $n$ edges). (See Figure 11). Suppose that when we measure an energy of this surface deformation, we obtain some quantity, $E^n$. What happens to the energy of this surface deformation, when we double the horizontal and vertical tessellation rates, so that now it a triangulated 2n-by-2n grid?

Let us first look at $E_{arap}^n$. The angles of the triangulated $S$ do not change, so the cotangents are unchanged. each vector $\mathbf{x}_v$ and $\mathbf{x}'_v$ multiplied by $1/2$. The resulting squared norm is multiplied by $1/4$. But the number of vertices along the fold doubles, so we obtain

$$E_{arap}^{2n} = \frac{1}{2} E_{arap}^n$$

In the limit, as the tessellation rate goes up, the discrete measured energy for a "fold" goes to zero. (The same behavior holds for the ARAP-Spoke-Rim energy). Thus we see that ARAP's does not resist bending in the limit of high tessellations.

Let us contrast this with the behavior of $E_{db}$ for this same folding deformation of $S$ to $S'$. Again let us see what happens when we double the tessellation rate. $H_v = 0$ so we can ignore that term. $\mathbf{x}'$ is multiplied by $1/2$. $\frac{\mathbf{x}'}{M_v}$ is multiplied by 2. $\|\frac{\mathbf{x}'}{M_v}\|^2$ is multiplied by 4. $M_v \|\frac{\mathbf{x}'}{M_v}\|^2$ is multiplied by 1. Again, the number of such vertices is doubled, and thus we obtain

$$E_{db}^{2n} = 2 E_{db}^n$$

At high tessellation rates, this blows up to infinity, which is what we would want for such infinite bending. Note that for this deformation, our stretching energy is always zero.

Indeed our computer experiments closely match this reasoning:

**Table 1:** *The energies between a flat sheet and a folded sheet*

| n | Triangles | Spoke | Spoke-Rim | Bending |
|---|---|---|---|---|
| 10 | 200 | 2343.1 | 4448.3 | 24.2 |
| 20 | 800 | 1171.5 | 2283.6 | 50.9 |
| 40 | 3200 | 585.7 | 1156.6 | 104.2 |
| 80 | 12800 | 292.8 | 582.1 | 210.9 |
| 160 | 51200 | 146.4 | 291.9 | 424.2 |
| 320 | 204800 | 73.2 | 146.2 | 850.9 |
| 640 | 819200 | 36.6 | 73.1 | 1704.2 |

**Case 2**

A similar analysis can be attempted for the mapping of a flat sheet $S$ to a cylinder $S'$ (see Figure 11). A very rough back-of-the-envelope calculation suggests that for this, case, we will find $E_{arap}^{2n} = \frac{1}{4} E_{arap}^n$ and $E_{db}^{2n} = E_{db}^n$. This behavior is born out by the computer experiments tabulated below. (Again, note that for this deformation, our stretching energy is always zero.)

**Table 2:** *The energies between a flat sheet and a cylinder*

| n | Triangles | Spoke | Spoke-Rim | Bending |
|---|---|---|---|---|
| 10 | 200 | 13460.2 | 25636.4 | 41.3 |
| 20 | 800 | 3663.6 | 7153.8 | 47.1 |
| 40 | 3200 | 947.3 | 1872.2 | 49.9 |
| 80 | 12800 | 240.3 | 477.8 | 51.3 |
| 160 | 51200 | 60.4 | 120.6 | 51.9 |
| 320 | 204800 | 15.1 | 30.2 | 52.3 |
| 640 | 819200 | 3.7 | 7.5 | 52.4 |

## C  More experiments

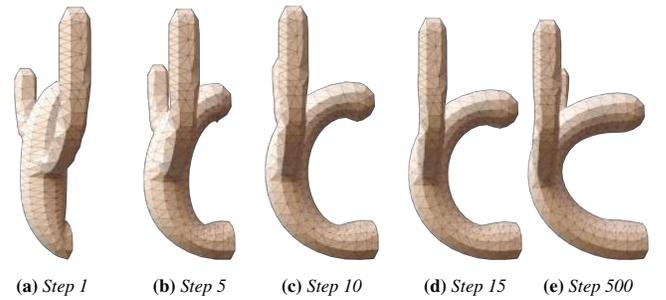

(a) *Step 1*  (b) *Step 5*  (c) *Step 10*  (d) *Step 15*  (e) *Step 500*

**Figure 12:** *Various numbers of iterations are shown.*

In figure 12, we show the local-global method converges fast, deformation result in the step 10 is almost the same with the step 500.

If table 3 , we compares the all kinds of deformations and different $\lambda$ in the hybrid method.

In table 4 , we show the different $\lambda$; In table 5, we show different constraints.

In table 8, we show the different $\lambda$ for hybrid methods.

**Table 3:** *Stretching, Bending, Hybrid with different λ*

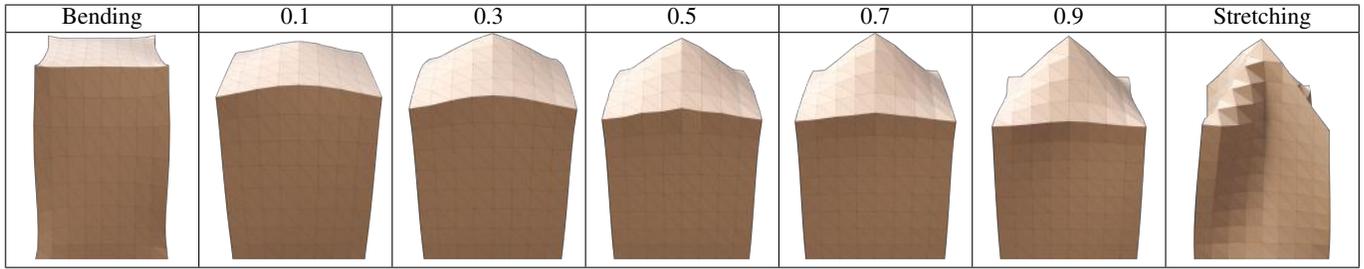

**Table 4:** *Stretching, Bending, Hybrid with different λ*

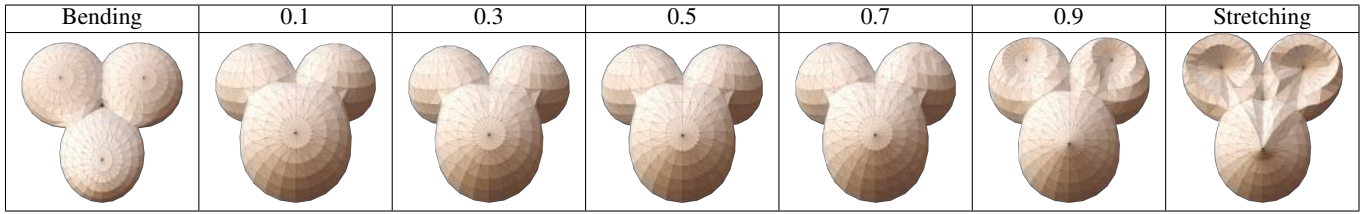

**Table 5:** *Stretching, Bending, Hybrid with different λ*

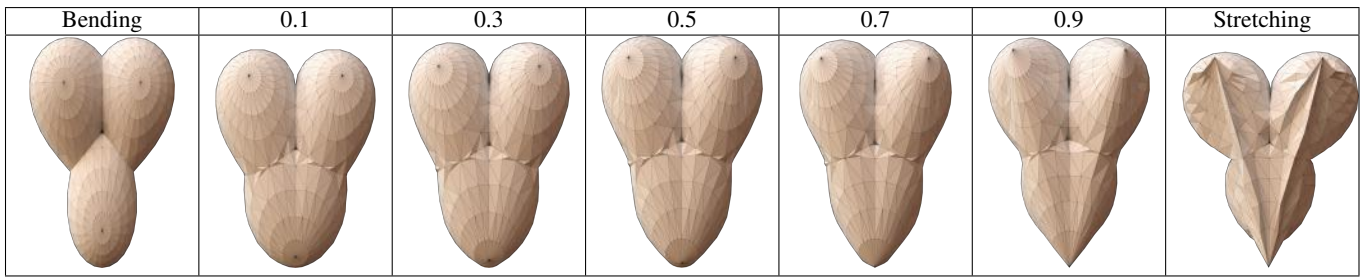

**Table 6:** *Stretching, Bending, Hybrid with different λ*

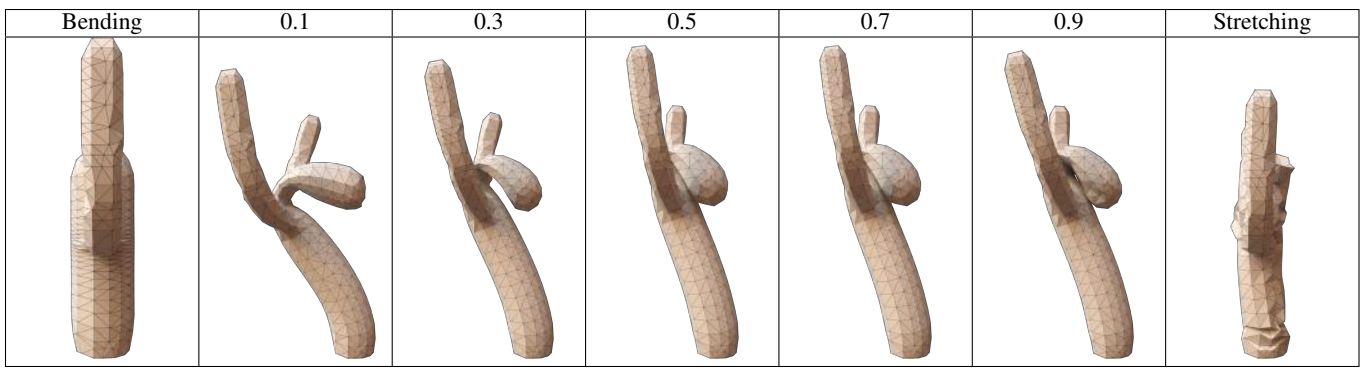

**Table 7:** *Stretching, Bending, Hybrid with different λ*

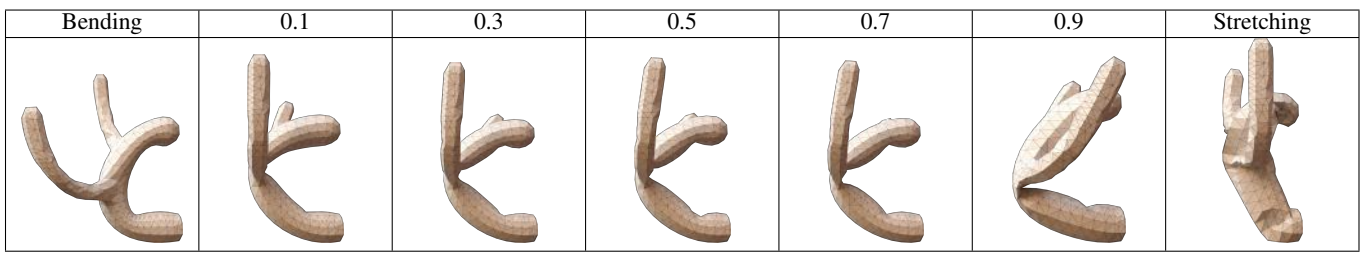

**Table 8:** *Stretching, Bending, Hybrid with different λ*

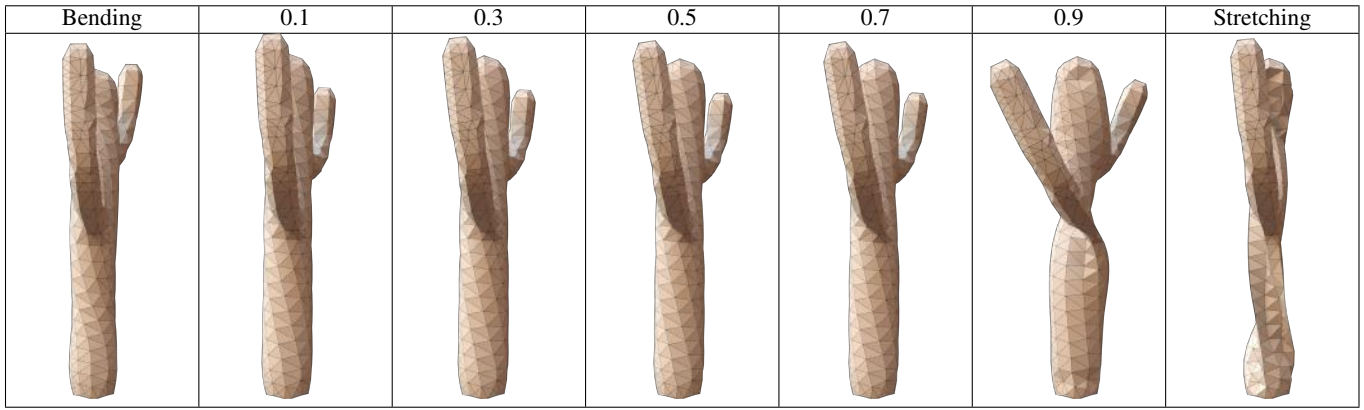

**Table 9:** *Stretching, Bending, Hybrid with different λ*

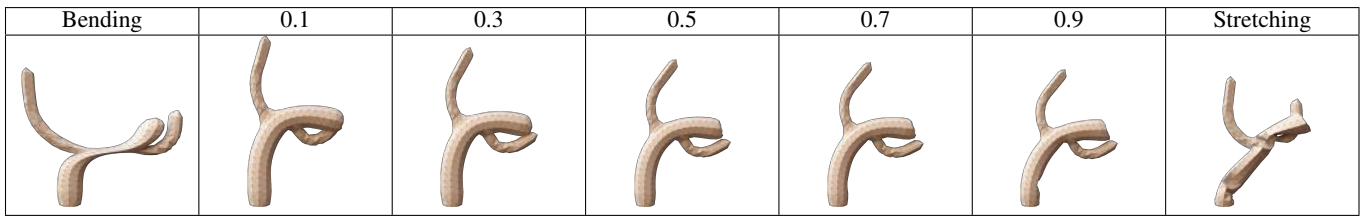

**Table 10:** *Stretching, Bending, Hybrid with different λ*

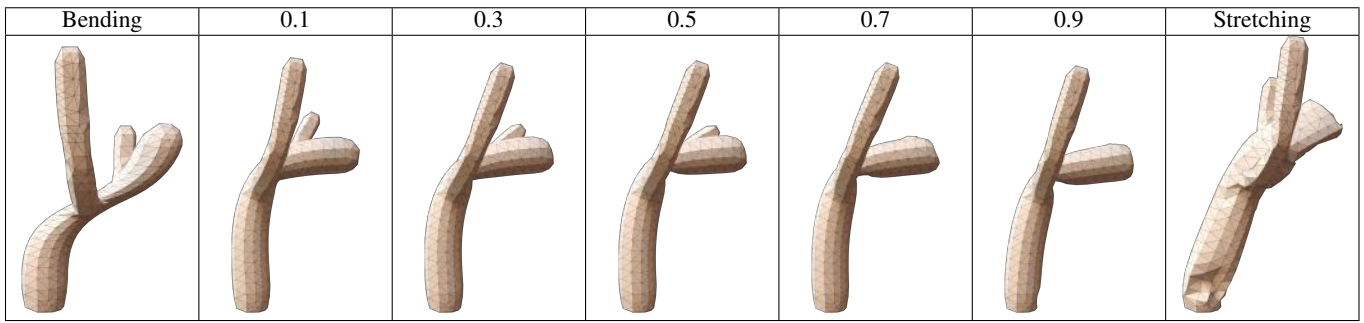

**Table 11:** *Hybrid with different λ*

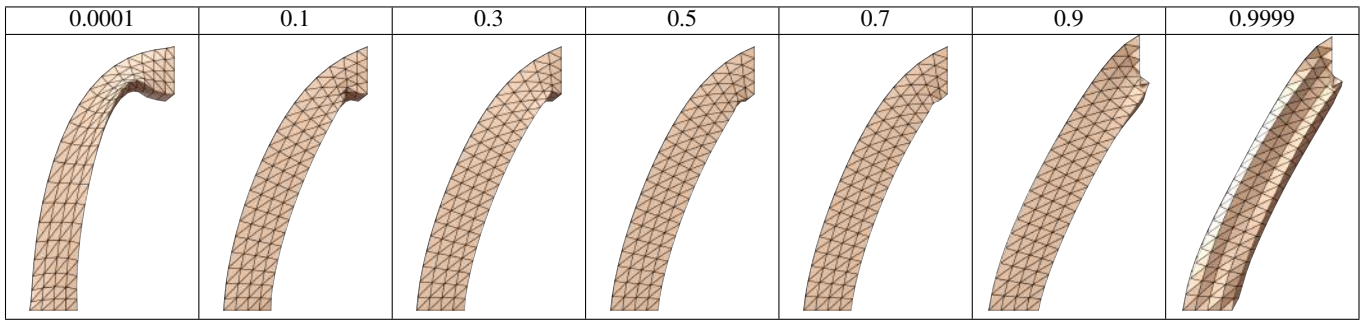

Table 12: *Hybrid with different* λ

| 0.0001 | 0.1 | 0.3 | 0.5 | 0.7 | 0.9 | 0.9999 |
|---|---|---|---|---|---|---|

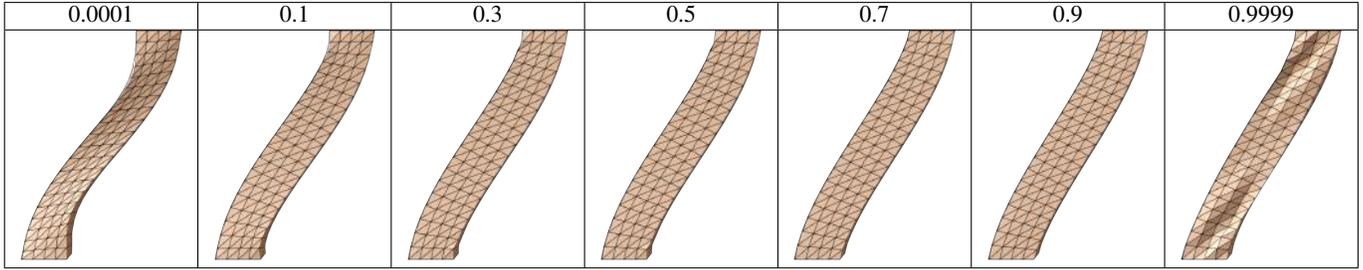

Table 13: *Hybrid with different* λ

| 0.0001 | 0.1 | 0.3 | 0.5 | 0.7 | 0.9 | 0.9999 |
|---|---|---|---|---|---|---|

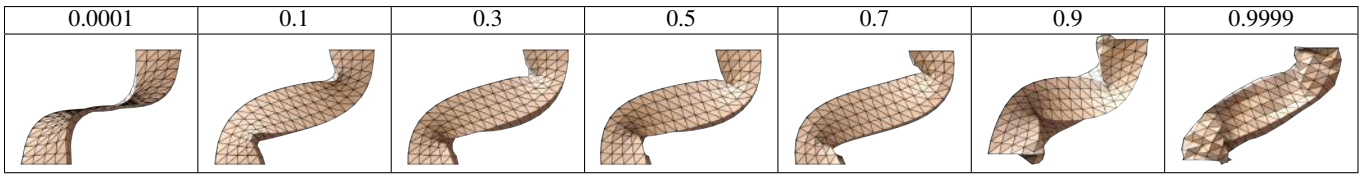

Table 14: *Hybrid with different* λ

| 0.0001 | 0.1 | 0.3 | 0.5 | 0.7 | 0.9 | 0.9999 |
|---|---|---|---|---|---|---|

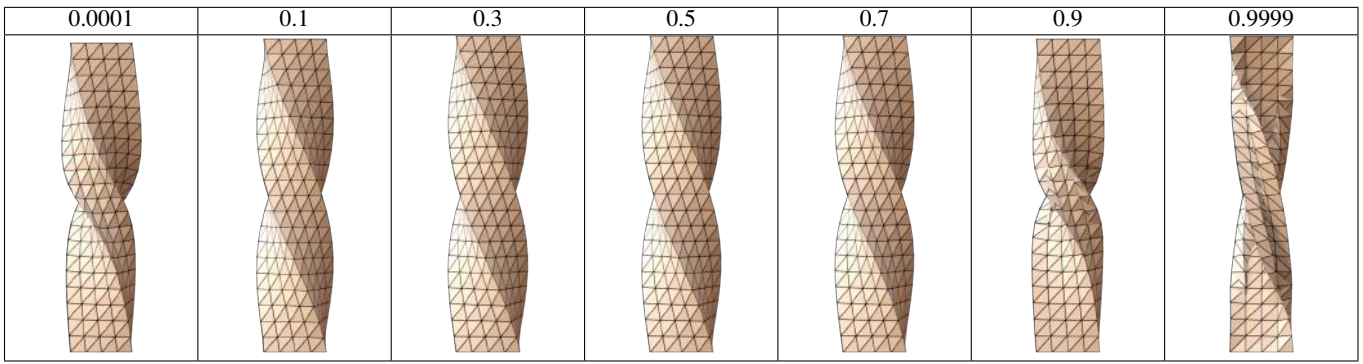

Table 15: *Hybrid with different* λ

| 0.0001 | 0.1 | 0.3 | 0.5 | 0.7 | 0.9 | 0.9999 |
|---|---|---|---|---|---|---|

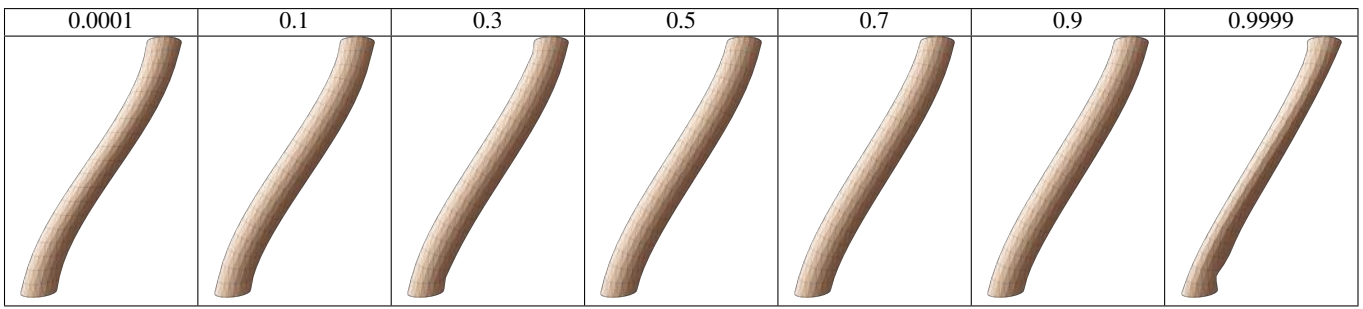

Table 16: *Hybrid with different* λ

| 0.0001 | 0.1 | 0.3 | 0.5 | 0.7 | 0.9 | 0.9999 |
|---|---|---|---|---|---|---|

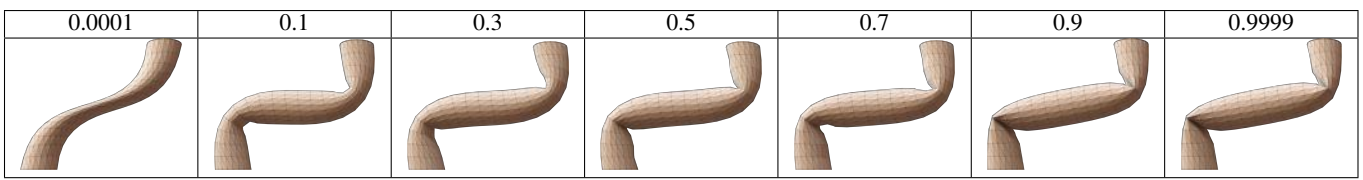

Table 17: *Hybrid with different* $\lambda$

| 0.0001 | 0.1 | 0.3 | 0.5 | 0.7 | 0.9 | 0.9999 |
|---|---|---|---|---|---|---|
| 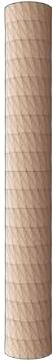 | 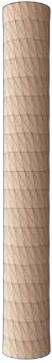 | 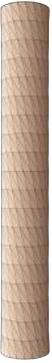 | 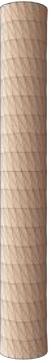 | 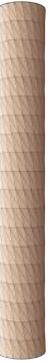 | 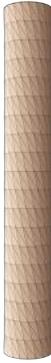 | 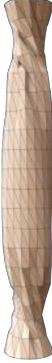 |